\documentclass[12pt]{article}
\usepackage[utf8]{inputenc}
\usepackage[english]{babel}
\usepackage[centertags]{amsmath}
\usepackage{amssymb}
\usepackage{bm}
\usepackage{amsbsy}
\usepackage{graphicx}
\usepackage{appendix}
\usepackage{mathrsfs}
\usepackage{epsfig}
\usepackage{color}
\usepackage{euscript}
\usepackage{ulem}
\usepackage{multirow}
\usepackage{cite}

\definecolor{dgreen}{rgb}{0,0.6,0}

\definecolor{darkblue}{rgb}{0., 0, 1}

\definecolor{purple}{rgb}{0.65,0.,0.78}

\definecolor{orange}{rgb}{0.89,0.42,0.05}

\usepackage{jheppubm}

\newcommand{\nn}{\nonumber}
\newcommand{\be}{\begin{equation}}
\newcommand{\ee}{\end{equation}}
\newcommand{\bea}{\begin{eqnarray}}
\newcommand{\eea}{\end{eqnarray}}

\newcommand{\fg}{\mathfrak{g}}

\newcommand{\fB}{\mathfrak{B}}

\newcommand{\cN}{{\cal N}}
\newcommand{\cK}{{\cal K}}

\numberwithin{equation}{section}

\title{Black Brane/Bose Gase Duality and \\ Third Law of Thermodynamics}

\author{Irina Aref'eva$^a$, Daniil Stepanenko$^a$ and Igor Volovich$^a$}

\affiliation{$^a$Steklov Mathematical Institute, Russian Academy of  Sciences, \\ Gubkina str. 8, 119991, Moscow, Russia}

\emailAdd{arefeva@mi-ras.ru, dstepanenko@mi-ras.ru, volovich@mi-ras.ru}

\abstract{

In the thermodynamics of black holes in asymptotically flat space, the third law of thermodynamics is violated, and entropy cannot be consistently modeled through conventional statistical mechanics. Notably, the third law of thermodynamics is violated for the Schwarzschild black hole, and its entropy can only be described using an unconventional model, such as a Bose gas in negative dimensions.

In contrast, for certain black brane solutions—such as Poincare AdS black branes, Lifshitz black branes, and anisotropic Lifshitz-type black branes—the third law is preserved, with entropy vanishing as temperature approaches zero. In this paper, we extend the previously established duality between black hole and Bose gas thermodynamics to black branes. Specifically, the Poincare black brane in $D$ spacetime dimensions corresponds to a non-relativistic Bose gas in $2(D-2)$ spatial dimensions. Furthermore, the duality between Lifshitz branes and Bose gases relates a Lifshitz brane with exponent $\alpha$ in $D$-dimensional spacetime to a Bose gas of quasi-particles with energy $k^\alpha$ in $D-2$ spatial dimensions.
}


\begin{document}

\maketitle

\section{Introduction}

There exists a remarkable analogy between black hole mechanics and thermodynamics \cite{Bardeen:1973gs,Bekenstein:1973ur}. However, the ordinary third law of thermodynamics is violated by asymptotically flat black hole solutions, such as the Schwarzschild, Reissner-Nordström, and Kerr solutions \cite{Israel,Wald,Arefeva:2023kpu,Arefeva:2023kwm}, and more references in \cite{Arefeva:2023kpu}. For these solutions (except in extremal cases), the entropy blows up at zero temperature.
\\

In this paper, we discuss the validity of the third law of thermodynamics for black brane solutions. This applies not only to the simplest black brane in AdS but also to non-relativistic black branes, such as Lifshitz branes \cite{Kachru:2008yh,Taylor:2008tg,Pang:2009ad,Balasubramanian:2009rx,Ayon-Beato:2010vyw}, and magnetic anisotropic black branes \cite{Arefeva:2016phb}, as well as p-brane solutions found in \cite{Klebanov:1996un,Arefeva:1997zrl}. For charged black branes, the entropy at zero temperature is bounded but not zero. In this case, there is an extremal limit corresponding to non-zero entropy at $T = 0$.
Having $S \neq 0$ at $T = 0$ is uncommon for a thermodynamic system and seems to imply a highly degenerate ground state \cite{Horowitz:2010nh}. This can be related to instability at low temperature when the system is coupled to other matter (e.g., a scalar field), where the black hole can have $S = 0$ at $T = 0$, as takes place for Lifshitz branes (see Sect.~\ref{sect:BHT} and Sect.~\ref{sec:Lt-BG}).
\\

Furthermore, we demonstrate a duality between these specific black brane solutions and the thermodynamics of a Bose gas \cite{LLV,huang,VZ,pitaevskii,pathria}, extending previously established results \cite{Arefeva:2023kpu,Arefeva:2023kwm,IAIV}. In particular, the Poincare black brane in $D$ spacetime dimensions is dual to a non-relativistic Bose gas in $2(D-2)$ spatial dimensions. This is in contrast to Schwarzschild black holes, which are dual to Bose gas models in negative dimensions. Physical quantities, such as free energy and entropy in negative dimensions, were interpreted in \cite{Arefeva:2023kpu,Arefeva:2023kwm} through analytical continuation.
\\

Moreover, the duality between Lifshitz branes and Bose gases associates a Lifshitz brane with an exponent $\alpha$ in $D$-dimensional spacetime to a Bose gas of quasi-particles with energy $k^\alpha$ (for a more precise expression for the quasi-particle energy see Eq.~\eqref{epsilon-s}) in positive $D-2$ spatial dimensions. Additionally, Lifshitz-type anisotropic branes in 5 dimensions with an anisotropy parameter $\nu$ are dual to a Bose gas with energy $k^\gamma$ in $d$ spatial dimensions, where $\gamma = d(2/\nu +1)$. 
\\

Lifshitz models have applications in condensed matter physics \cite{Erdmenger}, and Lifshitz-type models also appear in relativistic physics \cite{Arefeva:2014vjl,Arefeva:2014kyw}. Modeling them through the Bose gas framework opens new avenues to study these physically important modes.
\\

The paper is organized as follows. In Sect.~\ref{sect:Setup}, we present the main thermodynamic formulas for black holes and black branes (Sect.~\ref{sect:BH}) as well as the well-known formulas for a Bose gas (Sect.~\ref{sect:BGd}). In Sect.~\ref{sect:SBH-BG}, we review the previously established duality between \(D\)-dimensional Schwarzschild black holes and Bose gas models in negative dimensions, emphasizing the third law violation. 
From Sect.~\ref{sect:PBH-BG} through Sect.~\ref{sect:LAB-BG}, we extend this analysis to specific black brane solutions, including Lifshitz and anisotropic Lifshitz-type black branes. These brane solutions respect the third law of thermodynamics, allowing us to establish a duality between them and a standard Bose gas in positive dimensions. The corresponding calculations are detailed in Sect.~\ref{sect:PB-BG} through Sect.~\ref{sec:Lt-BG}.

\section{Setup}\label{sect:Setup}

\subsection{Black Hole and Black Brane Thermodynamics}\label{sect:BH}

Here, we consider the metric in a $D$-dimensional spacetime of the form
\bea
\label{metrigds2}
ds^2 = \fB^2(z) \left( -g(z) \cN(z) \, dt^2 + \sum_{i=1}^{D-2} \fg_i \, dx_i^2 + \frac{dz^2}{\cK(z) g(z)} \right),
\eea
where $t \in \mathbb{R}$, all transverse coordinates $x_i$ also belong to $\mathbb{R}$, and $z > 0$. The functions $\fB(z)$, $g(z)$, $\cN(z)$, and $\cK(z)$ are smooth and defined on appropriate domains, with the assumptions $\cN(z) > 0$ and $\cK(z) > 0$. The black hole horizon $z_h$ is defined as the root of the blackening function $g(z)$:
\bea
g(z_h) = 0.
\eea

The Hawking temperature can be found from the requirement that there is no conical singularity for the Euclidean version of the metric \eqref{metrigds2} near the horizon.
 Supposing that  $t$  varies from 0 to $\beta$,
we get the condition
\be
\frac12g'(z_h) \sqrt{\cK (z_h) \cN(z_h)}\beta=2\pi.\ee
Since $T=1/\beta$, we get
\be
T=\frac1{4\pi}g'(z_h)\sqrt{\cK (z_h) \cN(z_h)}.\ee
The  entropy is given by 

\bea\label{ent-gen}
S&=&\frac{V_{D-2}}{4 G_{D}}\left(\fB(z_h)\right)^{D-2}\prod_{i=1}^{D-2}\sqrt{\fg_i(z_h)},
\eea
where $V_{D-2}$ is the volume of the transverse directions.

\subsection{Bose Gas in $d$-Dimensional Space}
\label{sect:BGd}

Consider a Bose gas in a $d$-dimensional space \cite{LLV,huang,VZ,pitaevskii,pathria}, with the grand potential (free energy) given by
\bea
\label{2.2mm}
F_{BG} = \frac{1}{\beta} \sum_{\tiny{\begin{array}{c}
    k_i = 2\pi n_i/L \\
    i = 1, \ldots, d \\
    n_i = 1, 2, \ldots
\end{array}}}  \ln\Big(1 - e^{\beta\left(\mu - \sigma \, \varepsilon_\gamma(\vec{k})\right)}\Big),
\eea
where $\varepsilon_\gamma(\vec{k})$ is the energy of quasi-particles:
\bea
\label{epsilon-s}
 \varepsilon_\gamma(\vec{k}) = k^\gamma \equiv \left(\vec{k}^2\right)^{\gamma/2}, \qquad \vec{k} = \{k_1, \ldots, k_d\},
\eea
and $\sigma$ is a positive dimensional constant. The non-relativistic Bose gas corresponds to $\gamma = 2$.

For large $L$, the continuum version is
\bea 
\label{fs}
F_{BG}(\beta) = \frac{L^d}{(2\pi)^d \beta} \int \ln\left(1 - e^{\beta\left(\mu - \sigma \, \varepsilon_\gamma(\vec{k})\right)}\right) d^d k.
\eea 
Using spherical coordinates and integrating over spherical angles in \eqref{fs}, we obtain for $\mu = 0$:
\bea
\label{fsdalpha-angle}
F_{BG} = \frac{\Omega_{d-1}}{\beta} \left(\frac{L}{2\pi}\right)^d \int_{0}^{\infty} \ln\left(1 - e^{-\beta \, \sigma \, k^\gamma}\right) k^{d-1} \, dk,
\eea
where $\Omega_{d-1} = \frac{2 \pi^{d/2}}{\Gamma(d/2)}$, and $k^\gamma$ represents the exponent of the radial momenta as defined in \eqref{epsilon-s}.

By substituting the variable $\beta \, \sigma \, k^\gamma = x$ and using the representation
\bea 
\label{repr}
\int_{0}^{\infty} \ln\Big(1 - e^{-x}\Big) \, x^{s-1} \, dx = -\Gamma(s) \, \zeta(s+1),
\eea
valid for $\Re(s) > 1$, we obtain
\bea
 \label{dDIV}
 F_{BG} = -\left(\frac{L}{2\pi}\right)^d \frac{2 \pi^{d/2}}{d \, \Gamma(d/2)} \left(\frac{1}{\beta}\right)^{\frac{d}{\gamma} + 1} \left(\frac{1}{\sigma}\right)^{\frac{d}{\gamma}} \Gamma\left(\frac{d}{\gamma} + 1\right) \zeta\left(\frac{d}{\gamma} + 1\right).
\eea

The expression \eqref{dDIV} is obtained for $\Re \, d > \gamma$ (where $\Re$ denotes the real part), but $\zeta(s)$ can be analytically extended to the entire complex plane as a meromorphic function with a single simple pole at $s = 1$. Thus, we can use formula \eqref{dDIV} for complex values of $d$.

The entropy is given by
\bea
\nn
S_{BG} &=& -\frac{\partial F_{BG}(\beta, \gamma, \sigma, d, z)}{\partial T} = \beta^2 \frac{\partial F_{BG}}{\partial \beta}\\
&=& \left(1 + \frac{d}{\gamma}\right) \left(\frac{L}{2\pi}\right)^d \frac{2 \pi^{d/2}}{d \, \Gamma(d/2)} \left(\frac{1}{\beta \sigma}\right)^{\frac{d}{\gamma}} \Gamma\left(\frac{d}{\gamma} + 1\right) \zeta\left(\frac{d}{\gamma} + 1\right).
\label{SBGgamma}
\eea

\section{Black Hole and Bose Gas Duality in the Case of Third Law of Thermodynamics Violation}
\label{sect:SBH-BG}

\subsection{D-dimensional Schwarzschild Black Hole}
\label{sect:SBH}

The Euclidean action in $D$-dimensional space for the Schwarzschild black hole is:
\be
\label{act-d}
I_E = - \frac{1}{16 \pi G_{D}} \int_{\mathcal{M}} d^Dx \sqrt{g} R - \frac{1}{8 \pi G_{D}} \int_{\partial \mathcal{M}} d^{D-1}x \sqrt{h} K,
\ee
where the second term is the Gibbons-Hawking-York (GHY) term \cite{Gibbons:1976ue, York:1986it}, $h$ is the determinant of the induced metric on the boundary, and $K$ is the extrinsic curvature.

The Euclidean Schwarzschild metric is given by:
\be
\label{dsDSch}
ds^2 = \left(1 - \left(\frac{r_h}{r}\right)^{D-3}\right) d\tau^2 + \left(1 - \left(\frac{r_h}{r}\right)^{D-3}\right)^{-1} dr^2 + r^2 d{\bf \Omega}_{D-2}^2,
\ee
where $d{\bf \Omega}_{D-2}^2$ represents the metric on the $(D-2)$-sphere \cite{Myers:1986un}.

The horizon $r_h$ is related to the ADM mass by
\be
\label{ADM}
M = \frac{r_h^{D-3} (D-2) \Omega_{D-2}}{16\pi G_{D}},
\ee
where
\be
\label{OmegaD}
\Omega_{D-2} = \frac{2\pi^{(D-1)/2}}{\Gamma((D-1)/2)}
\ee
is the surface area of a unit sphere $S^{D-2}$.

The temperature of the Schwarzschild black hole is
\be
\label{TDSch}
T = \frac{D - 3}{4\pi r_h} = \frac{D - 3}{4\pi} \left( \frac{(D-2) \Omega_{D-2}}{16 \pi G_{D} M} \right)^{\frac{1}{D-3}}.
\ee
We observe that the Hawking temperature increases as $M$ decreases for $D > 3$, and the heat capacity $\frac{\partial M}{\partial T}$ is negative.

The entropy of the Schwarzschild black hole is
\be
\label{SDSch}
S = \frac{\Omega_{D-2} r_h^{D-2}}{4 \, G_{D}} = \frac{\Omega_{D-2}}{4 \, G_{D}} \left(\frac{D-3}{4\pi T}\right)^{D-2}.
\ee
We see that entropy increases as the Hawking temperature $T$ decreases, indicating a violation of the third law of thermodynamics.

In $D=4$ spacetime, the Schwarzschild black hole metric takes the form:
\be
\label{metric4}
ds^2 = -\left(1 - \frac{r_h}{r}\right) dt^2 + \frac{dr^2}{1 - \frac{r_h}{r}} + r^2 d{\bf \Omega}_{2}^2,
\ee
where $d{\bf \Omega}_{2}^2$ is the metric on the 2-dimensional unit sphere, $r_h = 2G_4 M$, and $M$ is the mass of the black hole.

The Hawking temperature and the Bekenstein-Hawking entropy are given by:
\be
T = \frac{1}{8 \pi G_4 M}, \qquad S = \frac{1}{16 \pi G_4 T^2}.
\ee
The entropy diverges as $T \to 0$, thus violating the third law of thermodynamics in Planck's formulation.

\subsection{Schwarzschild Black Hole/Bose Gas Duality}
\label{sect:SBH-BG-n}

This duality has been studied in detail in \cite{Arefeva:2023kpu, Arefeva:2023kwm}. Here, we briefly summarize the main result. By equating the entropy of a $D$-dimensional Schwarzschild black hole \eqref{SDSch} with the entropy \eqref{SBGgamma} of a $d$-dimensional Bose gas with energy \eqref{epsilon-s}, we obtain
\bea
\frac{\Omega_{D-2}}{4 \, G_{D}} \left(\frac{D-3}{4\pi T}\right)^{D-2} =
\left(1 + \frac{d}{\gamma}\right) \left(\frac{L}{2\pi}\right)^d \frac{2 \pi^{d/2}}{d \, \Gamma(d/2)} \left(\frac{T}{\sigma}\right)^{\frac{d}{\gamma}} \Gamma\left(\frac{d}{\gamma} + 1\right) \zeta\left(\frac{d}{\gamma} + 1\right),
\nn\\
\label{SDBG}
\eea
which leads to
\be
D - 2 = -\frac{d}{\gamma}.
\ee

This result implies that a positive $D - 2$ corresponds to a negative $d$. For negative $d$ and integer values of $1/\gamma$, singularities appear on the right-hand side of \eqref{SDBG} and one has to perform renormalizations, for further details see \cite{Arefeva:2023kwm}. The concept of negative dimension was also used in \cite{Volovich:2023vib} to compute the inflationary cosmological constant.

\section{Planar AdS  Black  Branes and Bose Gas}\label{sect:PBH-BG}
\subsection{Poincare Black Brane in AdS }\label{sect:PBH}

Note that for the Poincare-AdS black brane (i.e., the Schwarzschild-AdS black hole in the Poincare patch), also referred to as the planar AdS (PAds) black brane, the third law of thermodynamics holds.
Indeed, consider the action
\be
\label{PA}
S=\frac{1}{16\pi G_D}\int d^D x \sqrt{-g} \left[R-\Lambda\right].
\ee

The metric 
\be 
\label{Pm}
ds^2 = \frac{L^2}{z^2} \left( -g(z) \, dt^2 + d\vec{x}^2_{D-2} + \frac{dz^2}{g(z)} \right),
\ee
with the blackening function given by
\be 
g(z) = 1 - \left( \frac{z}{z_h} \right)^{D-1},
\ee
where $z_h$ is the black hole horizon, defined by $g(z_h) = 0$, solves the equations of motion derived from the action \eqref{PA}, for
\be
\Lambda = -\frac{(D-1)(D-2)}{2 L^2},
\ee
where $L$ is the AdS radius.

The Hawking temperature and the Bekenstein-Hawking entropy are \cite{Maldacena:2003nj}

\be T = \frac{D-1}{4 \pi z_h},\ee 
\be \label{SBB}
S \,= \,\frac{V_{D-2}}{4 G_D }\frac{L^{D-2}}{z_h^{D-2}}\,=\,
\frac{V_{D-2}}{4 G_D}\left(\frac{4 \pi  L T}{D-1}\right)^{D-2},
\ee
the volume  $V_{D-2}$ is determined by the area of the change of the coordinates  $x_1,... x_{D-2}$.

We see that $S \to 0$ as $T \to 0$, indicating that the third law of thermodynamics holds for the Poincare AdS black brane.

\subsection{PAdS Black Brane/Bose gas duality}\label{sect:PB-BG}

Let us now discuss the duality between the PAdS black brane and a Bose gas.  
The entropy of a Bose gas in a $d$-dimensional space is given by \eqref{SBGgamma}. Comparing this with the entropy of the black brane \eqref{SBB}, we see that both entropies have the same temperature dependence if
\be
\frac{d}{2} = D - 2.
\ee
This condition is satisfied in the special case where $D = d = 4$.

\section{Lifshitz  Black 
Brane}\label{sect:LBB-BG}
\subsection{Thermodynamics of Lifshitz  Black 
Hole}\label{sect:LBB}
Let us consider the  Lifshitz black brane solutions \cite{Kachru:2008yh,Taylor:2008tg,Pang:2009ad,Balasubramanian:2009rx}. The action that supports this brane solution can be taken in the form 
\be\label{MT}
S=\frac{1}{16\pi G_{D}}\int
d^{D}x\sqrt{-g}[R-\Lambda-\frac{1}{2}\partial_{\mu}\phi\partial^{\mu}\phi
-\frac{1}{4}e^{\lambda\phi}F_{\mu\nu}F^{\mu\nu}],
\ee
where $\Lambda$ is the cosmological constant and the matter fields
are a massless scalar and an abelian gauge field. 
The  metric in the form
\be
ds^{2}=L^{2}[-r^{2\alpha}f(r)dt^{2}+\frac{dr^{2}}{r^{2}f(r)}+r^{2}\sum\limits^{D-2}_{i=1}dx^{2}_{i}]
\ee
with matter fields 
\bea
\label{MTF}
F_{rt}= q\,e^{-\lambda\phi}r^{\alpha-D-3},\qquad
\phi= \pm\sqrt{2(\alpha-1)(D-2)}\log r\quad \mu!!!!!\eea
solves EOM for action \eqref{MT}
for $\alpha,L$  and $q$ related with $\lambda$ and $\Lambda$  as
\bea\label{MTq}
\alpha&= &1+\frac{2(D-2)}{\lambda^{2}},\\
L^2 &= &- \frac{(\alpha + D-3) (\alpha + D-2)}{\Lambda},\\
q^{2}&= &2L^2(\alpha-1)(\alpha+D-2).
 \eea
We assume that $\alpha \geq 1$.
The blackening function is
\begin{equation}
f(r)=1-\frac{r^{\alpha+D-2}_{h}}{r^{\alpha+D-2}},
\end{equation}
see \cite{Taylor:2008tg,Pang:2009ad} for details. Note that  $\alpha=1$ corresponds to $\phi =0$ and reproduces plane brane solution \eqref{Pm}. 

The temperature is
\be
\label{TMT}      T=\frac{(\alpha+D-2)}{4\pi  L}\,r^{\alpha}_h.
\ee
One can write the entropy as a function of
temperature
\be
\label{SMT}
    S_{BH}=\frac{V_{d}}{4G_{D}}\left(L\,r_h\right)^{D-2}=\frac{V_{d}L^{D-2}}{4G_{D}}\left(\frac{4\pi}{\alpha+D-2}L\,T\right )^{\frac{D-2}{\alpha}},
\ee
here $V_{D-2}$ is dimensionless.

\subsection{Lifshitz Black Brane/Bose Gas Duality} \label{sect:BHT}
Duality between Lifshitz black hole and Bose gas is given by the following formulae.
Equalizing the entropy of  the perfect Bose gas of quasi-particles with energy 
\eqref{epsilon-s} in $d$-dimensional space \cite{Arefeva:2023kpu,Arefeva:2023kwm} given by \eqref{SBGgamma} and entropy of Lifshitz black brane given by \eqref{SMT} we get an equality
\bea
\label{BGQP}
\left(\frac{L_{BG}}{2\pi^{1/2}}\right)^d\frac{\Gamma \left(\frac{d}{\gamma
   }+2\right) }{\Gamma(\frac{d}{2}+1)}\zeta \left(\frac{d
   }{\gamma }+1\right)\left(\frac{T}{\lambda
   }\right)^{\frac{d}{\gamma }}
    =\frac{V_{D-2}L^{D-2}}{4G_{D}}\left(\frac{4\pi L \,T}{\alpha+D-2}\right)^{\frac{D-2}{\alpha}}
   \label{BGL}
\eea
(here in \eqref{SBGgamma} $L\to L_{BG}$).
We see that if we take 
\be
d=D-2, \qquad \gamma =\alpha 
\ee
 and 
 we ensure that the equation  \eqref{BGL} is fulfilled  by adjusted the corresponding scale parameters $L$ and $L_{LQ}$. In this case we have to require that $\gamma>1$.

We can also suppose that $d$ and $D$ are related as 
\be\label{n}
d n =D-2,\ee
or
\be\label{m}
d  =(D-2 )m,\ee
where $n$ and $m$  are  natural numbers.
Since from \eqref{BGL} we get 
\be \label{d-gamm}\frac{d}{\gamma} =\frac{D-2}{\alpha},\ee
and relations \eqref{n} and \eqref{m} give us 
\bea
\label{nm}\alpha=n\gamma \quad \mbox{or}\quad  m\alpha=\gamma.\eea
The first solution in \eqref{nm}
allows us  to abandon the requirement $\gamma>1$. 

To summarize this section, we establish the duality between Lifshitz branes
 with exponent $\alpha$ in $D$-dimensional spacetime and Bose gases of quasi-particles with energy $(\vec k,\vec k)^{\alpha/2}$  in $D-2$ spatial dimensions. We can relax the condition   $\gamma=\alpha$ and consider a more general condition \eqref{d-gamm}.

\section{Anisotropic Lifshitz-type Black 
Brane}\label{sect:LAB-BG}
\subsection{Thermodynamics of Lifshitz-type  Black 
Brane}\label{sect:LAB}

The anisotropic Lifshitz-type black brane \cite{Arefeva:2014vjl, Arefeva:2016phb} is supported by the same action given in Eq.~\eqref{MT}.
The metric and matter fields for $D=5$ are given in \cite{Arefeva:2016phb} and Sect.~B.2 in \cite{Arefeva:2018hyo}:

\begin{equation} \label{mAGG}
ds^2=\frac{L^2}{z^2}\left(-g(z) \, dt^2 + dx_1^2 + \frac{dx_2^2 + dx_3^2}{(z/L)^{2/\nu - 2}} + \frac{dz^2}{g(z)}\right),
\end{equation}
with
\begin{align} \label{gAGG}
g(z) &= 1 - \left(\frac{z}{z_h}\right)^{2 + 2/\nu}, \quad 
F_{23} = - F_{32} = q, \quad \phi(z) = \frac{2 \sqrt{\nu - 1} \, \log(z/L)}{\nu}, \quad \nu \geq 1,
\end{align}
where the constants $\nu$, $\lambda$, $L$, $q$, and $\Lambda$ are related as
\begin{align}
\lambda &= -\frac{2}{\sqrt{\nu - 1}}, \label{lambda1} \\
\Lambda &= -\frac{2 (1 + \nu)(1 + 2\nu)}{L^2 \nu^2}, \label{lambda} \\
q^2 &= \frac{4 (\nu^2 - 1)}{L^2 \nu^2}. \label{q}
\end{align}
Thus, our solution is defined by the cosmological constant $\Lambda$ and the parameter $\lambda$, while other parameters, including $L$, are determined by Eqs.~\eqref{lambda1}--\eqref{q}.

The temperature and entropy are 
\be
\label{TAGG}
T=\frac{1+\nu}{\nu}\frac{1}{2 \pi }\frac1{z_h},
\ee

\be
\label{SAGG} S=\frac{V_{3}}{4 G_{5}}\left(\frac{L}{z_h}\right)^{2/\nu +1}=\frac{V_{3}}{4 G_{5}}\left(\frac{\nu}{1+\nu}2 \pi T\,L\right)^{2/\nu +1}.
\ee

\subsection{Lifshitz-type Black Hole/Bose Gas Duality}
\label{sec:Lt-BG}

Equating \eqref{BGQP} and \eqref{SAGG}, we obtain
\bea
\left(\frac{L_{BG}}{2\pi^{1/2}}\right)^d \frac{\Gamma \left(\frac{d}{\gamma} + 2\right)}{\Gamma\left(\frac{d}{2} + 1\right)} \zeta \left(\frac{d}{\gamma} + 1\right) \left(\frac{T}{\sigma}\right)^{\frac{d}{\gamma}} = \frac{V_{3}}{4 G_{5}} \left(\frac{\nu}{1 + \nu} 2 \pi T\, L\right)^{\frac{2}{\nu} + 1},
\eea
which leads to
\be
\label{d-gamma}
\frac{d}{\gamma} = \frac{2}{\nu} + 1.
\ee

Since we are free to choose any natural number as the dimension $d$ for the Bose gas, the given black brane model corresponds to a set of Bose gas models with $\gamma$ and $d$ satisfying the condition \eqref{d-gamma}.

\section{Conclusion}

The fact that black holes are dual to a Bose gas in negative dimensions explains the violation of the third law of thermodynamics by black holes. Indeed, the entropy of black holes corresponds to the entropy of a Bose gas in negative dimensions, which, according to Eq.~\eqref{SBGgamma}, depends on temperature through a negative power:
\begin{equation}
S_{BG} \sim T^{d/\gamma},
\end{equation}
where, for negative \( d < 0 \), we observe that \( S_{BG} \) diverges as \( T \to 0 \), leading to a violation of the third law of thermodynamics.
In light of Eq.~\eqref{SBGgamma}, this relationship can also be interpreted in the opposite direction: the inverse temperature dependence of black hole entropy suggests that a duality between black holes and a Bose gas is feasible only in negative dimensions.
\\

In this paper, we have established a duality between certain black branes and a Bose gas in standard positive dimensions. This result immediately implies the validity of the third law of thermodynamics in these cases.
\\

As for further directions, it would be interesting to generalize the duality between black branes and a Bose gas to include non-zero chemical potentials and to establish a connection between the Bose-Einstein condensate in a Bose gas and the phase transition of black branes.
\\

\section*{Acknowledgement}
We would like to thank A.~Golubtsova, A.~Hajilou, M. Khramtsov  and P. Slepov for useful discussions.
This work is supported by the Russian Science Foundation (24-11-00039, Steklov Mathematical Institute).


\begin{thebibliography}{99}
%\cite{Bardeen:1973gs,Bekenstein:1973ur}
\bibitem{Bardeen:1973gs}
J.~M.~Bardeen, B.~Carter and S.~W.~Hawking,
``The Four laws of black hole mechanics,''
Commun. Math. Phys. \textbf{31}, 161-170 (1973)
%doi:10.1007/BF01645742
%3304 citations counted in INSPIRE as of 17 Oct 2024
%\cite{Bekenstein:1973ur}
\bibitem{Bekenstein:1973ur}
J.~D.~Bekenstein,
``Black holes and entropy,''
Phys. Rev. D \textbf{7}, 2333-2346 (1973)
%doi:10.1103/PhysRevD.7.2333
%6720 citations counted in INSPIRE as of 17 Oct 2024

\bibitem{Israel}
W. Israel, Third Law of Black-Hole Dynamics. A Formulation and Proof,  Phys. Rev. Lett. {\bf 57}, 397 (1986).



\bibitem{Wald}
R. Wald,  "Nernst theorem" and black hole thermodynamics,  Phys. Rev. {\bf D56}, 6467 (1997)
%\cite{Arefeva:2023kpu,Arefeva:2023kwm}
\bibitem{Arefeva:2023kpu}%Arefeva:2023kwm
I.~Aref'eva and I.~Volovich,
``Violation of the third law of thermodynamics by black holes, Riemann zeta function and Bose gas in negative dimensions,''
Eur. Phys. J. Plus \textbf{139}, no.3, 300 (2024)
%doi:10.1140/epjp/s13360-024-05049-7
[arXiv:2304.04695 [hep-th]].
%\cite{Arefeva:2022guf}
\bibitem{Arefeva:2023kwm}
I.~Y.~Aref'eva and I.~V.~Volovich,
``Bose gas modeling of the Schwarzschild black hole thermodynamics,''
Theor. Math. Phys. \textbf{218}, no.2, 192-204 (2024)
%doi:10.1134/S0040577924020028
[arXiv:2305.19827 [hep-th]].
%2 citations counted in INSPIRE as of 16 Oct 2024



%Kachru:2008yh,Taylor:2008tg,Pang:2009ad,Balasubramanian:2009rx
\bibitem{Kachru:2008yh}
S.~Kachru, X.~Liu and M.~Mulligan,
``Gravity duals of Lifshitz-like fixed points,''
Phys. Rev. D \textbf{78}, 106005 (2008)
%doi:10.1103/PhysRevD.78.106005
[arXiv:0808.1725 [hep-th]].
%979 citations counted in INSPIRE as of 29 Sep 2024
\bibitem{Taylor:2008tg}
M.~Taylor,
``Non-relativistic holography,''
[arXiv:0812.0530 [hep-th]].

%\cite{Pang:2009ad}
\bibitem{Pang:2009ad}
D.~W.~Pang,
''A Note on Black Holes in Asymptotically Lifshitz Spacetime,''
Commun. Theor. Phys. \textbf{62}, 265-271 (2014)
%doi:10.1088/0253-6102/62/2/14
[arXiv:0905.2678 [hep-th]].
%73 citations counted in INSPIRE as of 30 Aug 2024
%\cite{Balasubramanian:2009rx}
\bibitem{Balasubramanian:2009rx}
K.~Balasubramanian and J.~McGreevy,
``An Analytic Lifshitz black hole,''
Phys. Rev. D \textbf{80}, 104039 (2009)
%doi:10.1103/PhysRevD.80.104039
[arXiv:0909.0263 [hep-th]].
%164 citations counted in INSPIRE as of 29 Sep 2024
%\cite{Taylor:2015glc}

\bibitem{Ayon-Beato:2010vyw}
E.~Ayon-Beato, A.~Garbarz, G.~Giribet and M.~Hassaine,
``Analytic Lifshitz black holes in higher dimensions,''
JHEP \textbf{04}, 030 (2010)
%doi:10.1007/JHEP04(2010)030
[arXiv:1001.2361 [hep-th]].
%126 citations counted in INSPIRE as of 11 Oct 2024

%\cite{Arefeva:2016phb}
\bibitem{Arefeva:2016phb}
I.~Y.~Aref'eva, A.~A.~Golubtsova and E.~Gourgoulhon,
``Analytic black branes in Lifshitz-like backgrounds and thermalization,''
JHEP \textbf{09} (2016), 142
%doi:10.1007/JHEP09(2016)142
[arXiv:1601.06046 [hep-th]].

%\cite{Horowitz:2010nh}

\bibitem{Klebanov:1996un}
I.~R.~Klebanov and A.~A.~Tseytlin,
``Entropy of near extremal black p-branes,''
Nucl. Phys. B \textbf{475} (1996), 164-178
 % [arXiv:hep-th/9604089 [hep-th]].

%\cite{Arefeva:1997zrl}
\bibitem{Arefeva:1997zrl}
I.~Y.~Aref'eva, M.~G.~Ivanov and I.~V.~Volovich,
``Nonextremal intersecting p-branes in various dimensions,''
Phys. Lett. B \textbf{406}, 44-48 (1997)
%doi:10.1016/S0370-2693(97)00630-8
[arXiv:hep-th/9702079 [hep-th]].
%63 citations counted in INSPIRE as of 08 Oct 2024
\bibitem{Horowitz:2010nh}
G.~T.~Horowitz,
``Surprising Connections Between General Relativity and Condensed Matter,''
Class. Quant. Grav. \textbf{28}, 114008 (2011)
doi:10.1088/0264-9381/28/11/114008
[arXiv:1010.2784 [gr-qc]].
%26 citations counted in INSPIRE as of 17 Oct 2024
\bibitem{Arefeva:2022guf}
I.~Aref'eva and I.~Volovich,
``Complete Evaporation of Black Holes and Page Curves,''
Symmetry \textbf{15}, no.1, 170 (2023)
%doi:10.3390/sym15010170
[arXiv:2202.00548 [hep-th]].
%12 citations counted in INSPIRE as of 21 Oct 2024


\bibitem{LLV} %,huang,VZ,pitaevskii,pathria
L. D. Landau, E. M. Lifshitz  Statistical Physics: Volume 5.  Elsevier, 2013,  v. 5.

\bibitem{huang}
K. Huang, 
\textit{Statistical Mechanics}, 
2nd ed., Wiley, 1987.

\bibitem{VZ} V.A. Zagrebnov and J.-B. Bru. "The Bogoliubov model of weakly imperfect Bose gas." Physics Reports 350.5-6 (2001) 291-434.

\bibitem{pitaevskii}
L. P. Pitaevskii and S. Stringari, 
\textit{Bose-Einstein Condensation}, 
Clarendon Press, 2003.

\bibitem{pathria}
R. K. Pathria and P. D. Beale, 
\textit{Statistical Mechanics}, 
3rd ed., Elsevier, 2011.


\bibitem{IAIV} I.~Y.~Aref'eva and I.~V.~Volovich,
``Black Hole Thermodynamics and Bose Gas with Non-zero Chemical Potential", in preparation.\\
I.~Y.~Aref'eva,  "Bose Gas/de Sitter duality," talk at Quantum gravity and Cosmology 2024, China, 
July 1-5, 2024




\bibitem{Erdmenger} M.~Ammon and J.~Erdmenger, "Gauge/gravity duality: Foundations and applications",
https://doi.org/10.1017/CBO9780511846373
%\cite{Arefeva:2014kyw}
\bibitem{Arefeva:2014vjl}
I.~Y.~Aref'eva and A.~A.~Golubtsova,
``Shock waves in Lifshitz-like spacetimes,''
JHEP \textbf{04}, 011 (2015)
%doi:10.1007/JHEP04(2015)011
[arXiv:1410.4595 [hep-th]].
%37 citations counted in INSPIRE as of 08 Oct 2024

\bibitem{Arefeva:2014kyw}
I.~Y.~Aref'eva,
``Holographic approach to quark\textendash{}gluon plasma in heavy ion collisions,''
Phys. Usp. \textbf{57}, 527-555 (2014)
doi:10.3367/UFNe.0184.201406a.0569
%76 citations counted in INSPIRE as of 08 Oct 2024
%\cite{Arefeva:2018hyo}
\bibitem{Arefeva:2018hyo}
I.~Aref'eva and K.~Rannu,
``Holographic Anisotropic Background with Confinement-Deconfinement Phase Transition,''
JHEP \textbf{05}, 206 (2018)
%doi:10.1007/JHEP05(2018)206
[arXiv:1802.05652 [hep-th]].
%99 citations counted in INSPIRE as of 31 Oct 2024
\bibitem{Volovich:2023vib}
I.~Volovich,
``Cosmological Constant and Maximum of Entropy for de Sitter Space,''
[arXiv:2308.11377 [hep-th]].
%1 citations counted in INSPIRE as of 16 Oct 2024



%\cite{Gibbons:1976ue,York:1986it}
\bibitem{Gibbons:1976ue}
G.~W.~Gibbons and S.~W.~Hawking,
``Action Integrals and Partition Functions in Quantum Gravity,''
Phys. Rev. D \textbf{15}, 2752-2756 (1977)
%doi:10.1103/PhysRevD.15.2752
%3470 citations counted in INSPIRE as of 16 Oct 2024


\bibitem{York:1986it}
J.~W.~York, Jr.,
``Black hole thermodynamics and the Euclidean Einstein action,''
Phys. Rev. D \textbf{33}, 2092-2099 (1986)
%doi:10.1103/PhysRevD.33.2092
%480 citations counted in INSPIRE as of 16 Oct 2024 

\bibitem{Myers:1986un}
R.~C.~Myers and M.~J.~Perry,
``Black Holes in Higher Dimensional spacetimes,''
Annals Phys. \textbf{172}, 304 (1986)
%doi:10.1016/0003-4916(86)90186-7
%2070 citations counted in INSPIRE as of 15 Oct 2024


\bibitem{Maldacena:2003nj}
J.~M.~Maldacena,
``TASI 2003 lectures on AdS/CFT,''
[arXiv:hep-th/0309246 [hep-th]].
%207 citations counted in INSPIRE as of 08 Oct 2024

\end{thebibliography}
\end{document}